\documentclass[journal]{IEEEtran}
\usepackage[letterpaper, left=1in, right=1in, bottom=1in, top=0.75in]{geometry}
\usepackage{graphicx}
\usepackage{nicefrac}
\usepackage{multirow}
\usepackage{url}
\usepackage{hyperref}
\usepackage{bigstrut}
\usepackage{amsmath,amssymb,amsfonts}
\ifCLASSOPTIONcompsoc
  \usepackage[nocompress]{cite}
\else
  \usepackage{cite}
\begin{document}
\title{Airborne Urban Microcells with Grasping End Effectors: A Game Changer for 6G  Networks?}
\author{Vasilis~Friderikos% <-this % stops a space
\IEEEcompsocitemizethanks{\IEEEcompsocthanksitem V. Friderikos is with the Department
of Engineering, Centre for Telecommunications Research, King's College London, London, Strand, WC2R 2LS, UK\protect\\
E-mail: \href{vasilis.friderikos@kcl.ac.uk}{vasilis.friderikos@kcl.ac.uk}}% <-this % stops a space
}

\markboth{}%
{Shell \MakeLowercase{\textit{et al.}}: Bare Advanced Demo of IEEEtran.cls for IEEE Computer Society Journals}

\IEEEtitleabstractindextext{%
\begin{abstract}
Airborne (or flying) base stations (ABSs) embedded on drones or unmanned aerial vehicles (UAVs) can be deemed as a central element of  envisioned 6G cellular networks where significant cell densification with mmWave/Terahertz communications will be part of the ecosystem. Nonetheless, one of the key challenges facing the deployment of ABSs is the inherent limited available energy of the drone, which limits the hovering time for serving ground users to the orders of minutes.  This impediment deteriorate the performance of the UAV-enabled cellular network and hinders wide adoption and use of the technology. 
In this paper, we propose robotic airborne base stations (RABSs) with grasping capabilities to increase the serving time of ground users by multiple orders of magnitude compared to nominal hovering based operation. More specifically, to perform the grasping task, the RABS is equipped with a versatile, albeit, general purpose gripper manipulator. Depending on the type of the gripper  RABS can provide service in the range of hours, compared to minutes of hovering based ABSs. In theory it is possible that grasping can be energy neutral, hence the time of service can be bounded by the communications energy consumption.  To illustrate the case, energy consumption comparison between hovering and grasping is performed in order to reveal the significant benefits of the proposed approach. Finally, overarching challenges, design considerations for RABS, and future avenues of research are outlined to realize the full potential of the proposed robotic aerial base stations.

\end{abstract}

}

\maketitle
\thispagestyle{empty}

% is not selected <OR> if conference mode is selected - because compsoc
% conference papers position the abstract like regular (non-compsoc)
\IEEEdisplaynontitleabstractindextext

% \fi
\IEEEpeerreviewmaketitle

\ifCLASSOPTIONcompsoc
\IEEEraisesectionheading{\section{Introduction}\label{sec:introduction}}
\else
\section{Introduction}
\label{sec:introduction}
\fi

\IEEEPARstart{D}{ue} to their highly versatile characteristics,  drones and/or UAVs operating as airborne base stations (ABSs) are expected to play a central role in future 6G wireless communications. Drones have an inherent flexible and robust three dimensional maneuverability, allowing for rapid and on-demand deployment of ABSs, whilst providing high quality line-of-sight (LoS) air-to-ground (A2G) communication links and hence, being able to provide ultra high capacity data offloading in hotspot areas and/or increase the coverage area of terrestrial base stations. 
%The issue of airborne Base Station (ABS) deployment is an important factor that affect the communication performance of UAV-aided wireless networks and the overall energy efficiency of those networks. Even though some deployment positions can be manually selected according to the population density [131], the increasing dynamics, variable propagation characteristics, complex physical surroundings, and even the climates drive the researchers and operators to consider more efﬁcient and automatic strategies
To this end, the use of Unmanned Aerial Vehicles (UAVs) assisted 6G communications in the form of  airborne base stations (ABSs)  has the potential to significantly increase terrestrial wireless network operations in a vast variety of different use cases such as hot spot data coverage, data collection for Internet of Things (IoTs) and various different machine-type communications. In those settings, UAV assisted wireless networks can boost network efficiency and flexibility by providing seamless on-demand high capacity connectivity for elastic traffic in hot spot areas as well as also in areas beyond the coverage area of a terrestrial base station (BS).

5G networks have been designed to provide peak aggregate data rates at the range of 20Gbps, with an average user experience rate of 120Mbps. However, in 6G networks aggregate rates are envisioned to reach 1Tbps with  1Gbps support per user in order to enable applications such as fully immersive augmented reality and holographic communications. To achieve that new higher frequency bands will be required. For example, the U.S. Federal Communications Commission (FCC) has opened frequency bands from 95 Gigahertz (GHz) to 3 THz for testing and verification regarding their potential for providing such new services in 6G networks.
\begin{figure}[htb]
	\centering
	\includegraphics[width=0.3 \textwidth]{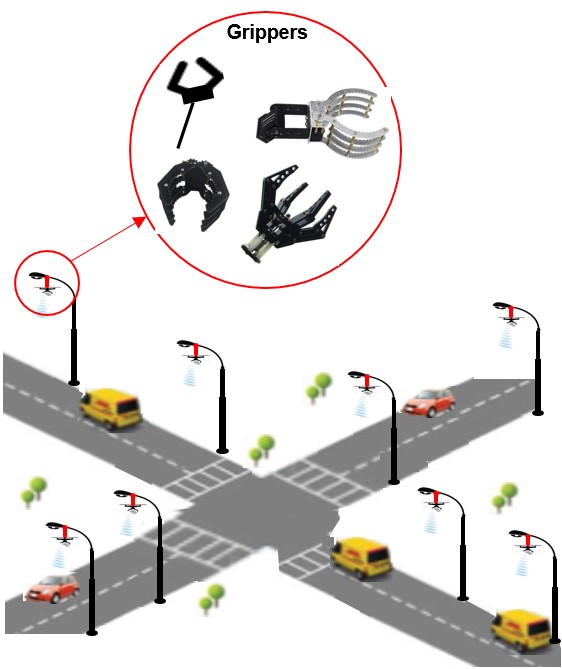}
	\caption{Illustrated use case of airborne 6G microcells with grasping end effectors (RABS). }
	\label{fig:toy}
\end{figure}
\\
Therefore, in the envisioned 6G ecosystem  high level of cell densification will inevitably have to take place and the use of Terahertz spectrum will require close proximity between 6G base stations and the end users. The stance to be taken hereafter is that robotic airborne base stations (RABS) with grasping capabilities can be considered as a highly suitable technology that can efficiently fulfil the above aims for 6G networks. The idea of utilizing RABS for, inter alia, cell densification is shown in figure \ref{fig:toy}. As can be seen from the figure, airborne base stations with grasping capabilities attach to roadside lamppost units and with an energy neutral gripper they can be deployed to provide 6G services for many hours. Given the importance of both mmWave and (sub)THz bands in 6G systems RABS could fullfil the niche role of ultra-high capacity cells in an energy efficient manner. This use case is inline with the investigations in \cite{rappaport}  where 6G base stations are assumed  at heights of 4
m above the ground, and the authors explicitly mention that is a similar height to lampposts.

%It is important to note however that the majority of the research work on UAV with grasping capabilities has been focused on the UAV being able to grasp objects of various shapes, textures, weights, and sizes rather than to anchor the UAV in a specific location. This requires a re-thinking of the manipulators which also relate also to the specification of the object where the RABS will be anchored.

For miniaturized UAVs the idea has already been proven feasible. The work in \cite{Perching} details a grasping mechanism  which is build on top of a quadcopter CrazyFlie  2.0\footnote{https://nettigo.eu/products/quadcopter-crazyflie-2} drone (note that this drone weight just 27gr).  The proposed gripper has two fingers that could either clip each side of an objects to utilize the friction forces for perching or encompass the object by forming a closed shape. In essence, RABS could be considered as massively scaled-up versions of such miniaturized drones, however significant more functionalities need to be embedded, as  detailed in the sequel, in order to allow for a fully autonomous operation.

There are also some auxiliary, albeit, significantly important benefits of RABS compared to nominal hovering based operation of airborne base stations. The first is that RABS ease dependency on weather conditions, allowing  for a \textbf{rainproof} operation. Even though hovering during rainy conditions is in theory possible, in reality electronics might not be able to shield enough to operate the task of hovering in rainy conditions. Furthermore, drone thrusting might be significantly compromised and to put that in context it is worthwhile noting that in 2020 in UK there were  170 days in which 1 mm or more of rain fell\footnote{www.statista.com/statistics/610677/annual-raindays-uk/}. RABS could be deployed proactively based on short term weather predictions and hence minimize such hazards. Also, when ambient temperature is below zero degrees Celsius hovering might be prohibited;  this might be true even for temperatures higher than zero degrees when there is significant levels of moisture in the atmosphere in which case the high speed of rotary wings might reduce the temperature and create a built-up of ice.
Furthermore, RABS are also \textbf{windproof} since hovering under windy conditions requires not only increased levels of energy consumption but might be infeasible. Currently, there are no accepted specific regulations regarding flying in windy conditions. For example, the UK Civil Aviation Authority mentions that ''Do not fly if the weather could affect your flight''\footnote{https://register-drones.caa.co.uk/drone-code/making-every-flight-safe} without providing any quantification about this recommendation as with respect to the underlying weather conditions. To put that in context (and without considering wind gusts and shear) drones might experience difficulty in flying/hovering at wind speeds that exceed 15Kt whereas the average wind speed in UK has been reported at 9.2kt for 2020\footnote{www.statista.com/statistics/322785/average-wind-speed-in-the-united-kingdom-uk/}. However, RABS could be proactively deployed using short term weather forecasting to again minimize such events.

\begin{figure}[htb]
	\centering
	\includegraphics[width=0.49 \textwidth]{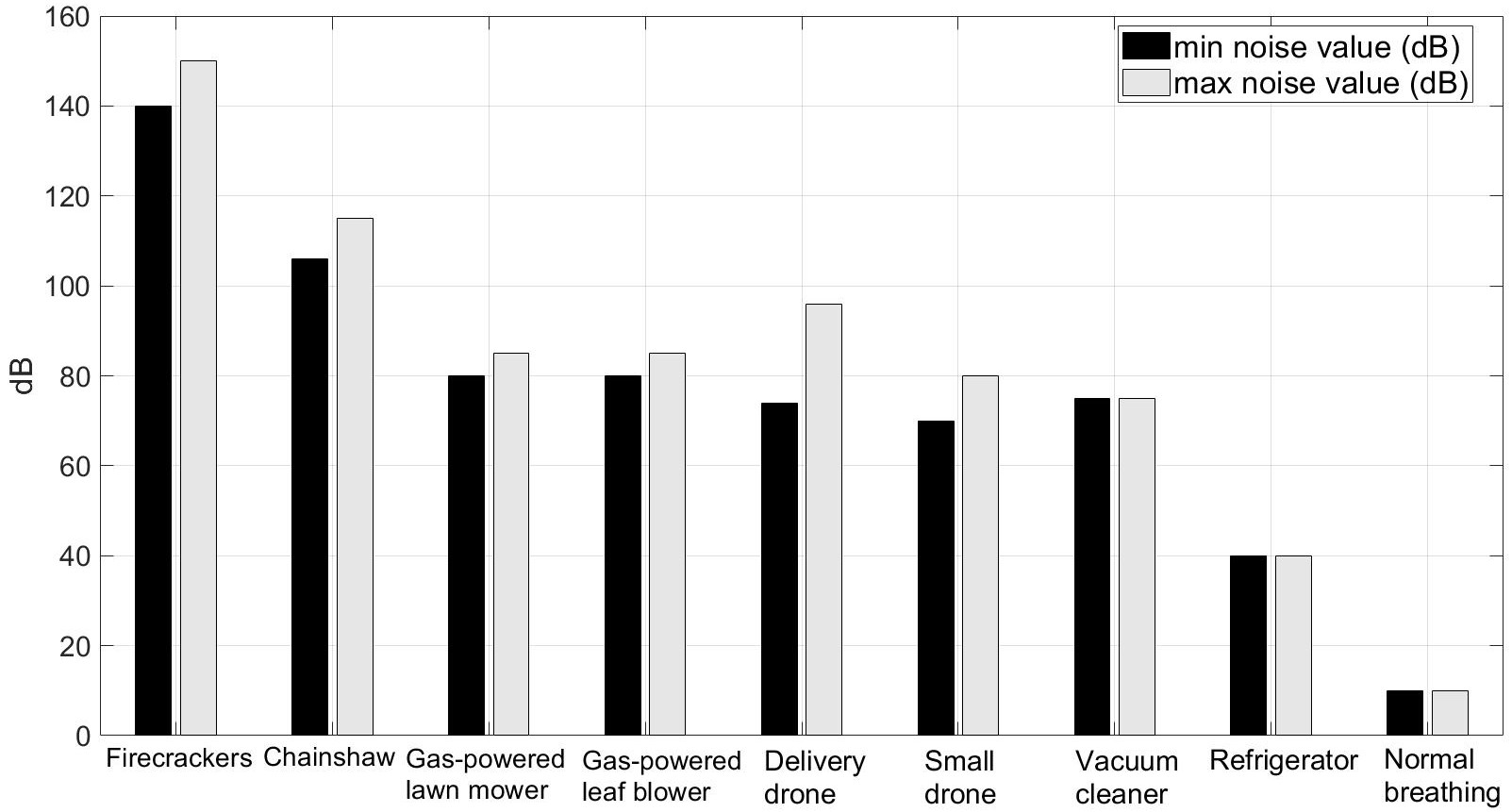}
	\caption{Noise level generated by drones in comparison with other common everyday activities. }
	\label{fig:noise}
\end{figure}

Furthermore, another key benefit of RABS is that they can provide  zero levels of  \textbf{noise pollution} by grasping into different objects. This is because  RABS can switch-off their rotors instead of hovering.% for boosting wireless network capacity. 
%This is of significant importance since  rotary wing drones are very noisy, many of them operating close the limit of acceptable noise, being 85dBA. 
Rotary wing drones when flying or hovering operate very close to the so-called  Acceptable-Noise-Level (ANL) which is at 85dB. To put that in context, Fig. \ref{fig:noise} provides a vis-a-vis comparison of the noise level emitted by drones compared to other typical every day activities\footnote{Data available from https://datawrapper.dwcdn.net/SzpE2/4}. Important to note that the limits for urban ambient noise are at 45dB during the night and at 55dB during the day \cite{Noise}. %Also, theoretically there is a distance–altitude relationship of a 6dB reduction with a doubling of the distance for rotary wing drones \cite{Noise2}.
Free-field sound decay rate is at approximately 6dB with a doubling of the distance. 
Therefore, for high capacity airborne cells utilizing mmWave communications at the height of interest can be deemed as an important element which, surprisingly, hasn't yet been thoroughly  considered. However, the proposed RABS can effectively eliminate noise pollution whilst serving ground users.
%Commercial drones, such as the Vanguard, are able to remain at 60dB even within 10m range.

\section{RABS System Architecture}
Airborne Base Stations (ABSs) that allow hovering to provide network coverage on a specific location falls under the multi rotor UAVs. These UAVs are versatile, economic and easy to manufacture. Broadly, the different categories of multi rotor UAVs range from  the tricopters, which have 3 rotors, to the octocopter with 8 rotors.

In addition to the 6G wireless communication components RABS should also be embedded with some further hardware and software components in order to allow autonomous operation; flying and grasping at different locations. 
%\subsection{Additional Suitable Use Cases for RABS}
%\subsubsection{RABS as Fog Computing Nodes}
%ABSs have been recently proposed as fog computing nodes due to their inherent high mobility and flexible deployment which can allow the provision of storage and computing resources close to the end users \cite{fog}, \cite{fog2}. The benefit is that UAV based fog computing nodes can be placed in locations with low demand and/or to augment existing network infrastructure without the need to place fixed servers in these locations. However, these works considered that the UAVs that act as fog computing nodes need to hover at specific locations (which are optimized) and as a result their use can be deemed as severely constrained due to the limited onboard battery resource. Hence, in this case these fog computing nodes can provide computing and storage support for very limited amount of time. On the other hand, RABS   
\subsection{Types of Grippers}
Grasping is one of the fundamental procedures performed by robotic manipulators. Therefore, the area of robotic grasping received significant research attention over the recent years and different types of robotic grippers have been taxonomized \cite{review_grippers}. However, for the grasping application in RABS there are in general two types of grippers that can be utilized; the electric grippers and the electro-magnetic grippers (whilst being possible to create hybrid designs). Electric grippers rely on electric motors that respond to precise input from a controller. The RABS control unit orchestrate the gripper with instructions in order to accomplish the grasping  task.  Electromagnetic based grasping \cite{electromagnet} technologies on the other hand due to their versatility can also be used especially in conjunction with a claw shaped gripper since typically the object where the RABS will grasp  will be metal  (i.e., a lamppost).
\begin{figure}[htb]
	\centering
	\includegraphics[width=0.3 \textwidth]{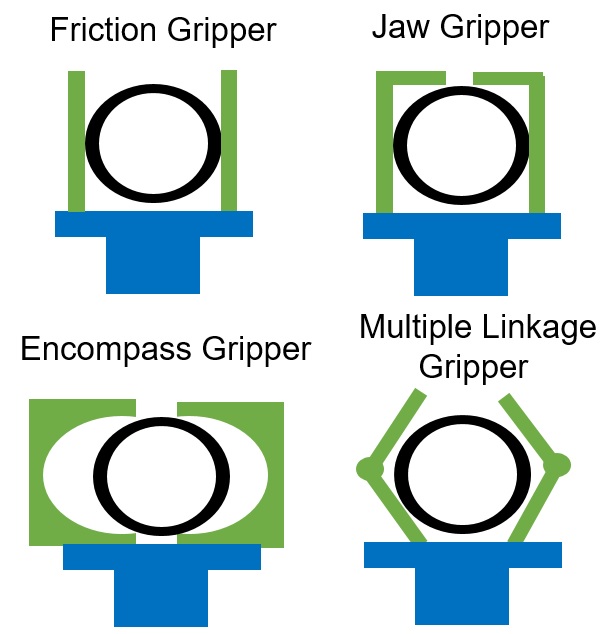}
	\caption{Different possible gripper configuration for the RABS. }
	\label{fig:grippers}
\end{figure}
The following simple calculations can reveal the required force in order to hang the RABS to lamppost in the worst case scenario which would be using a friction gripper. In that case the required force $F$ to hold the airborne base station would be, 
\begin{equation}
F > \dfrac{m (g+\alpha)}{\mu} \cdot \epsilon
\end{equation}
where $m$ is the mass of the RABS, $g$ is the gravitational constant, $\alpha$ is the acceleration which is assumed as zero, $\mu$ the static friction coefficient between the finger attachment and the point of attachment  and $\epsilon$ is safety factor which is typical equal to 2. Since typical values of the friction coefficient range between 0.1 to 0.2 the force requirement from the gripper to hold the typical RABS of 5 Kgr should range from 490N to 980N. However, note that an encompass or a jaw type of gripper (retention gripper as it is also commonly called) could allow RABS hanging from a lamppost in an almost energy neutral manner; since the gripper end effector will be able to withhold the weight of the RABS.
\subsection{Autonomous grasping} 
In order to perform grasping of the RABS at a specific object in an autonomous manner real-time target object detection is required. There are already a number of successfully integration efforts of vision based approaches for autonomous and real-time grasping in UAVs \cite{vision1}, \cite{vision2}. 
With the integration of deep learning-based algorithms the overall performance can be significantly achieved, however those schemes require significant  computational complexity which entail energy consumption. However, for the use case 
of RABS the actual communication and potentially MEC capabilities can be used in order to perform the autonomous grasping functionalities. Alternatively, these functionalities can be offloaded to an edge cloud and can be provided as a service to different RABS. 
\subsection{Communications Energy Consumption}
For airborne base stations acting as small cells we can assume an efficient design that reduces to the minimum the baseband signal processing, and the overall power consumption of the transceivers and their associated radio frequency hardware components 
For comparison reasons we will be assuming some nominal communication system parameters. Assuming a total communication bandwidth  being equal to 1 MHz,  then nominal  transmit power can reach up to 20 dBm, and the reference Signal to Noise Ratio (SNR) can be considered as being equal 52.5 dB. Then, the overall communication-related power consumption of the UAV would aggregate to approximately $Pc = 5 $ Watts \cite{energy}, \cite{coms_energy}. Also,  a drone base stations prototype from Nokia\footnote{www.nokia.com/about-us/news/releases/2016/10/03/f-cell-technology-from-nokia-bell-labs-revolutionizes-small-cell-deployment-by-cutting-wires-costs-and-time} (F-cell) has been designed as energy-efﬁcient small cell and the overall communication hardware components consume less than 15 Watts. Hence, the range from 5 to 15 W can effectively approximate the communication energy requirements of aerial base stations. In the numerical investigations hereafter we will be assuming the upper limit of this range.

\subsection{Energy Consumption for Flying and Hovering}
The energy consumption of typical rotary wing UAVs consists of two main components, namely, the propulsion and the acceleration/deceleration part.  The propulsion power consumption as a result of the flying speed in watts (W) can be approximately expressed as follows, \cite{energy}, \cite{energy2}, 
\begin{equation}
\begin{aligned} 
\label{f1}
P_{T}=&P_{0} \left(1+\frac{3V^2}{U_{tip}^2}\right) \\& +P_{i}\left(\sqrt{1+\frac{V^4}{4v_{0}^{4}}}-\frac{V^2}{2v_{0}^{2}}\right)^{1/2}+\frac{1}{2}d_{0} \rho \hat{s}AV^{3}\\
\end{aligned} 
\end{equation}

In the above expression
 $P_{0}$ and $P_{i}$ are  constants representing the blade profile power and the induced power when in hovering, respectively. Furthermore, $U_{tip}$ denotes the tip speed of the
rotor blade, $v_0$ is the mean rotor induced velocity in
hover, $d_0$ is the fuselage drag ratio.
Note that there are two types of UAV energy propulsion consumption; the $flying$ and $hovering$ mode energy consumption. Specifically, $P_{T}$ as expressed in Eqs. (\ref{f1}) denotes the $flying$ mode energy. However, we are interested in the $hovering$ mode energy which is defined as the energy when the UAV hovers at a specific location to serve ground users. The hovering energy consumption ($P_{H}$) can be derived by inserting $V=0$ into Eqs. (\ref{f1}). In this case the $hovering$ energy consumption is $P_{H} = P_O + P_i$\footnote{$\delta$: profile drag coefficient, $\rho$: air density in Kg per cubic meters, $s$: rotor solidity, $A$: rotor disc area in square meters, $\Omega$: blade angular velocity in radians, $R$: rotor radius in meters, $\kappa$: coefficient factor, $m$: total mass of the UAV in Kg, $g$: gravitational constant in meters/square$\cdot$sec. For a more detailed treatment on power consumption and aeromechanical aspects of UAVs the interested reader could follow the following textbook \cite{energy_book}. }, where,  
\begin{equation}
P_O = \frac{\delta}{8}\rho s A \Omega^3 R^3 
\end{equation}
\begin{equation}
P_i = (1+\kappa) \dfrac{(m g)^{3/2} }{\sqrt{2 \rho A}} 
\end{equation}

\subsection{Energy Consumption for Grippers}
Electromagnetic solenoid based grippers can be deemed as suitable for attaching to ferromagnetic surfaces such as lampposts. A typical lightweight industrial electromagnet is capable of lifting up to 40 kg of payload and hence can provide the necessary magnetic hold force for grasping. Such a gripper can also be implemented as a small array of electromagnets. Taking as an example the electromagnet mounted on an UAV in \cite{electromagnet} requires a constant 12 V DC supply for grasping and consumes 10 Joules/sec in full operation whilst being able to lift at maximum 6Kg; the electromagnet weights 400 gr. If we consider two 5V electromagnets of 25 Kg holding force each, translate to a usable weight hold of up to 5Kg assuming a 10 fold safety factor, which can represent the total weight of a RABS. The two electromagnets will draw 1.2A at 5 V DC. The table \ref{tab:solenoid} below provides a synopsis of three different solenoids with varying capabilities.
\begin{table}[!htbp]
%\begin{table}[!htbp]
\centering
\normalsize
\caption{Summary of Solenoid based Grippers}
\label{tab:solenoid}
\begin{tabular}{ |p{3cm}||p{1.2cm}|p{1.2cm}|p{1.2cm}|  }
 %\hline
 %\multicolumn{4}{|c|}{Solenoid based Grippers} \\
 \hline
 Characteristics & Type I & Type II & Type III\\
 \hline
 Holding force (Kg)    & 30    & 35 &   60\\
 Voltage DC (V) &   12  & 24   & 24\\
 Max. current (mA)  & 0.340 & 400 &  670 \\
 Power cons. (W) &   4  & 10 & 15 \\
 Weight (gr) & 193  & 186 & 346\\
 \hline
\end{tabular}
\end{table}
%%%%%%%%%%%%%%%
\begin{table}[!htbp]
%\begin{table}[!htbp]
\centering
\normalsize
\caption{Characteristics of Typical Drone Batteries}
\label{tab:batteries}
\begin{tabular}{ |p{2cm}||p{1.2cm}|p{1.2cm}|p{1.2cm}| p{1.2cm}| }
 %\hline
 %\multicolumn{4}{|c|}{Solenoid based Grippers} \\
 \hline
 Description & Capacity (mAh) & Voltage (V)  & Battery Type & Weight (g) \\
 \hline
 Zappers SG4   & 6100    & 15.2 &   LiPo & 424 \\
 GiFi Power &   8050  & 14.8   & LiPo & 650 \\
 Tattu  & 9000 & 14.8 &  LiPo & 810 \\
 Tattu &  10000  & 14.8 & LiPo & 935 \\
 Gens Ace & 11000  & 14.8 & LiPo & 963\\
 \hline
\end{tabular}
\end{table}
%\subsection{Energy Consumption Communications}
%An efficient design of a energy-efﬁcient aerial BS acting as a small cell that reduces to the minimum the baseband signal processing, and the power consumption of the transceivers and their associated radio frequency hardware components to consume in the range of 5 to 15 Watts as an absolute maximum value. 

%it is insigniﬁcant compared to the power drawn by the UAV’s onboard computer (60 W) and motors (1000 W) from the 8000 mAh ( 178 Wh) LiPo Battery. 
%%%%%%%%%%%%%%%%%%%%%
\subsection{Additional Suitable Use Cases for RABS}
\subsubsection{RABS as Fog Computing Nodes}
ABSs have been recently proposed as fog computing nodes due to their inherent high mobility and flexible deployment which can allow the provision of storage and computing resources close to the end users \cite{fog}, \cite{fog2}. The benefit is that UAV based fog computing nodes can be placed in locations with low demand and/or to augment existing network infrastructure without the need to place fixed servers in these locations. However, these works considered that the UAVs that act as fog computing nodes need to hover at specific locations (which are optimized) and as a result their use can be deemed as severely constrained due to the limited onboard battery resource. Hence, in this case these fog computing nodes can provide computing and storage support for very limited amount of time. On the other hand, RABS %\subsubsection{RABS as High Capacity Relay Nodes}
\begin{figure}[htb]
	\centering
	\includegraphics[width=0.48 \textwidth]{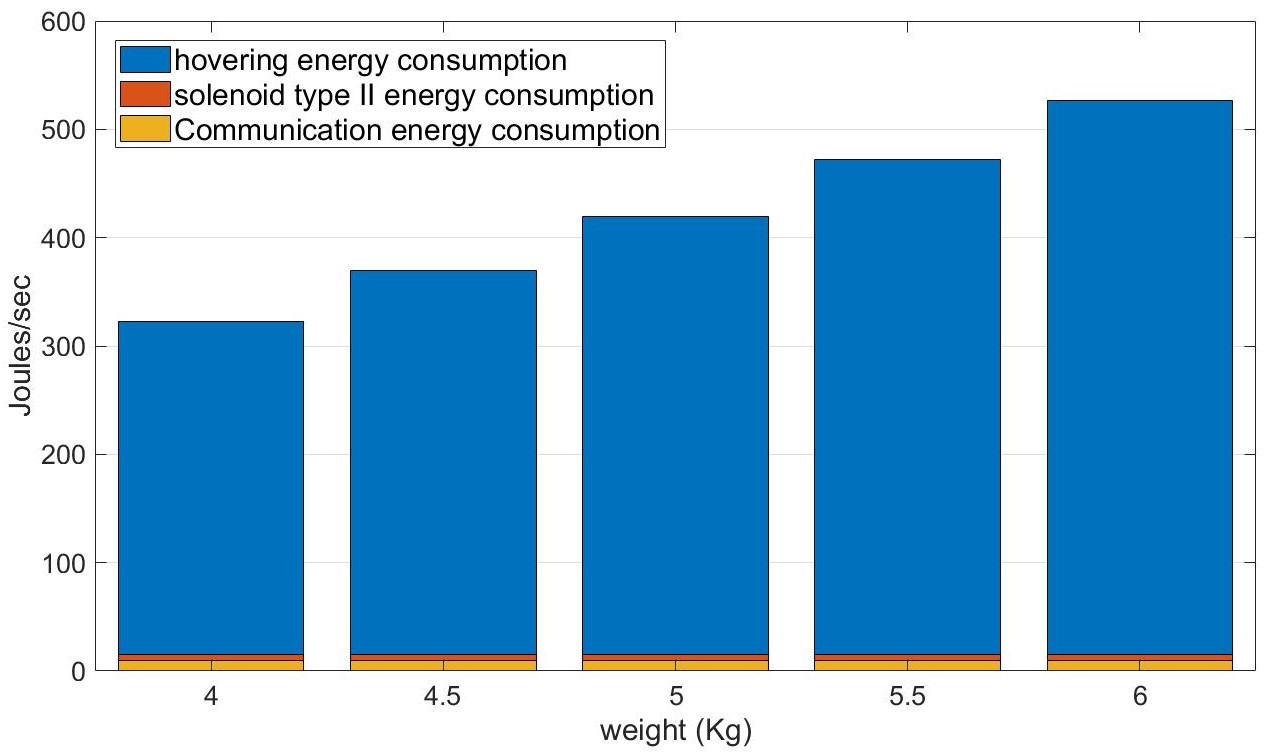}
	\caption{Comparison on energy consumption (Joules/sec) for hovering, grasping (solenoid Type II) and communications. }
	\label{fig:energy_comp}
\end{figure}
\begin{figure}[htb]
	\centering
	\includegraphics[width=0.48 \textwidth]{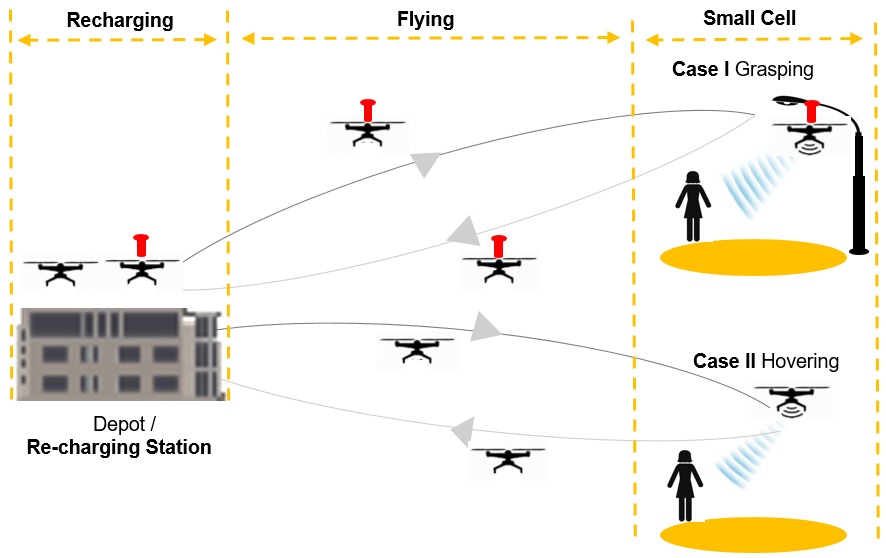}
	\caption{Use case scenario for comparing RABS with nominal hovering based ABSs. }
	\label{fig:usecase}
\end{figure}
\begin{table*}[t]
\caption{Comparison Between Nominal Airborne Base Stations (ABS) and Robotic Airborne Base Stations (RABS) with Grasping Capabilities}
\label{Tab:Comparison}
\centering
\begin{tabular}{|c|c|c|c|c|c|}
\hline 
D=800m, UAV weight 4Kg & ABS & RABS (+0.4Kg) & RABS (+0.8Kg) &  RABS (+1.2Kg) &  RABS (+1.6Kg) \\ [0.1cm]
\hline
Flying Energy (KJ) & 23.91 & 25.58 & 27.32 & 29.13 & 31.02 \\ [0.1cm]
Communications Energy (J/sec) &15 & 15 & 15 & 15 &15\\ [0.1cm]
Grasping Energy (J/sec) & 0 & 15 & 15 & 15 & 15\\ [0.1cm]
Hovering Energy (J/sec) & 323.05 & 0 & 0 & 0 & 0\\ [0.1cm]
Available time to serve (min) & \textbf{4.88} & \textbf{171.2} & \textbf{33.67} & \textbf{23.6} & \textbf{13.1}\\ [0.1cm] 
\hline
\end{tabular}
\end{table*}

\section{Indicative Performance Gains: Simulations and Discussion}
In this section indicative performance comparisons are detailed to shed further light on the potential gains of utilizing high capacity robotic airborne base stations. We consider a 6100 mAh battery at 15.2 V inline with typical values used by drones as shown in Table \ref{tab:batteries}.  Fig. ~\ref{fig:energy_comp} provides a comparison between the energy consumption between hovering, grasping and communications for different overall weight of the drone. As can be seen from the figure, the hovering energy consumption dominates the overall energy consumption of the micro base station. To illustrate the potential gains of RABS we consider a use case scenario as shown in Fig. \ref{fig:usecase}. The airborne base station departs from a depot that act as the re-charging station and after covering a distance $D$ by flying reach the hot spot area. We will be assuming a covering distance, $D$, equal to 800m which is equivalent to covering a 2$Km^2$ (to put that in perspective, the City of London $-$ London's historic and financial centre $-$ is approximately 2.9$Km^2$). The comparison is on the actual service time between hovering and grasping, and we don't  assume  an energy-neutral grasping function but one that consumes 15 J/Sec. However, weight difference due to the gripper end effector between RABS and a nominal ABSs which provide service by hovering is taken into account. Table ~\ref{Tab:Comparison} provides a comparison between the nominal hovering based ABS operation and the proposed robotic based ABS (RABS) with grasping capabilities. As can be seen from the Table the flying energy consumption is higher for RABS due to the extra weight of the grasping end effector. However, the net available time to serve is significantly increased as shown in the last row; RABS with the lightest grasping end effector of 0.4Kg  can prolong end users serving time by more than 35 times  compared to nominal hovering based operation. And, if we assume an energy neutral gripper serving time can reach 342.4 min, representing an almost 6 hours availability of acting as a 6G small cell before having to return to depot for recharging. %Such extensive increase on the actual serving time of airborne base stations can be considered as a key factor that can enable advanced 6G service provision.
\subsection{Research Challenges}
For all its aforementioned benefits RABS pose  a number of research challenges before creating commercial-grade  airborne microcells to integrate into a mobile network. One of the key research challenges is the design of grasping manipulators for RABS that will enable  energy neutral operation and being customized to typical lamppost settings or suitable places for grasping. Suitable auxiliary miniaturized cameras, depth sensors, or tactile sensors to assist autonomous grasping need to be investigated. Side information such known candidate location for gasping as well as prior knowledge of object geometry can greatly ease the requirements on grasping. Since RABS grasping to different surfaces and objects requires a number of different components and algorithms the process can be considered to be offered as a service, i.e, Grasping as a Service (GaaS). In this case, different advanced algorithms, including computer vision and AI techniques need to be developed that will run on  the edge cloud nodes to assist RABS for successful attachment to a specific object in a fully autonomous manner.
%\begin{figure}[htb]
%	\centering
%	\includegraphics[width=0.48 \textwidth]{comparison_case.jpg}
%	\caption{Use case scenario for comparing RABS with nominal hovering based ABSs. }
%	\label{fig:usecase}
%\end{figure}
%\begin{figure}[htb]
%	\centering
%	\includegraphics[width=0.48 \textwidth]{plot_energy_final.jpg}
%	\caption{Comparison on energy consumption (Joules/sec) for hovering, grasping (solenoid Type II) and communications. }
%	\label{fig:energy_comp}
%\end{figure}
% TABLE
%
%\begin{table*}[t]
%\caption{Comparison Between Nominal Airborne Base Stations (ABS) and Robotic Airborne Base Stations (RABS) with Grasping Capabilities}
%\label{Tab:Comparison}
%\centering
%\begin{tabular}{|c|c|c|c|c|c|}
%\hline 
%D=800m, UAV weight 4Kg & ABS & RABS (+0.4Kg) & RABS (+0.8Kg) &  RABS (+1.2Kg) &  RABS (+1.6Kg) \\ [0.1cm]
%\hline
%Flying Energy (KJ) & 23.91 & 25.58 & 27.32 & 29.13 & 31.02 \\ [0.1cm]
%Communications Energy (J/sec) &15 & 15 & 15 & 15 &15\\ [0.1cm]
%Grasping Energy (J/sec) & 0 & 15 & 15 & 15 & 15\\ [0.1cm]
%Hovering Energy (J/sec) & 323.05 & 0 & 0 & 0 & 0\\ [0.1cm]
%Available time to serve (min) & \textbf{4.88} & \textbf{171.2} & \textbf{33.67} & \textbf{23.6} & \textbf{13.1}\\ [0.1cm] 
%\hline
%\end{tabular}
%\end{table*}

\section{Conclusions}
This paper articulate on the potential of  airborne base stations with grasping end effectors to augment cellular networks capacity as 6G urban microcells. The proposed approach aims to tackle the key limitations of currently proposed UAV based base stations which relates to their limited duration of service support due to the high energy consumption for hovering. %To this end, we have explored the use of robotic airborne base stations (RABS) with grasping end effectors to tackle the hovering energy consumption problem. 
We have discussed the key benefits and different RABS configurations and analyze the potential gains in terms of extending the serving time compared to the nominal case which relates to hovering when serving ground users. Analysis
 showed that RABSs systems can have massive  advantages compared to nominal flying base stations, and can propel efficient 6G network densification especially for the use of emerging and future high frequency bands especially those in mmWave and (sub)Terahertz spectrum that, inherently, require close  proximity between the base station and the end-users.  
 %provide an irrefutable technology for 6G networks to provide massive capacity gains in a very dynamic and versatile manner.
Finally, and towards the vision 6G robotic base stations we discussed some open ended challenges and research problems that requires further investigation to enable fully autonomous 6G urban microcell deployment and operation. 
%and a synergy between the wireless communications and robotics community in order to design suitable RABSs for different urban and/or rural environments.

\ifCLASSOPTIONcaptionsoff
  \newpage
\fi

\end{document}